\documentclass[aps, pre, twocolumn, a4paper, floatfix]{revtex4-1}
\usepackage{graphicx}
\usepackage{amsmath}
\usepackage{placeins}
\usepackage{color}
\usepackage{hyperref}

\begin{document}

\title{Spontaneous centralization of control in a network of company ownerships}
\author{Sebastian M.\ Krause}
\email{sebastian.krause@itp.uni-bremen.de}
\author{Tiago P. Peixoto}
\email{tiago@itp.uni-bremen.de}
\author{Stefan Bornholdt}
\email{bornholdt@itp.uni-bremen.de}
\affiliation{Institut f\"ur Theoretische Physik, Universit\"at Bremen, D-28359 Bremen, Germany}
\begin{abstract}
We introduce a model for the adaptive evolution of a network of company
ownerships.  In a recent work it has been shown that the empirical
global network of corporate control is marked by a central, tightly
connected ``core'' made of a small number of large companies which
control a significant part of the global economy. Here we show how a
simple, adaptive ``rich get richer'' dynamics can account for this
characteristic, which incorporates the increased buying power of more
influential companies, and in turn results in even higher control. We
conclude that this kind of centralized structure can emerge without it
being an explicit goal of these companies, or as a result of a
well-organized strategy.
\end{abstract}
\maketitle

\section{Introduction}

The worldwide network of company ownership provides crucial information
for the systemic analysis of the world
economy~\cite{schweitzer_networks_2009,farmer_complex_2012}. A complete
understanding of its properties and how they are formed has a wide range
of potential applications, including assessment and evasion of systemic
risk~\cite{battiston_debtrank_2012}, collusion and antitrust regulation
\cite{gulati_strategicnetworks_2000,gilo_collusion_2006}, market
monitoring~\cite{diamond_delegated_1984,chirinko_germanbanks_2006}, and
strategic investment~\cite{teece_cooperation_1992}. Recently, Vitali et
al~\cite{vitali_network_2011} inferred the network structure of global
corporate control, using the Orbis 2007 marketing
database~\footnote{\url{http://www.bvdinfo.com/products/company-information/international/orbis}}.
Analyzing its structure, they found a tightly connected ``core'' made of
a small number of large companies (mostly financial institutions) which
control a significant part of the global economy. A central question
which arises is what is the dominant mechanism behind this
centralization of control. The answer is not obvious, since the decision
of firms to buy other firms can be driven by diverse goals: Banks act as
financial intermediaries doing monitoring for uninformed investors
\cite{diamond_delegated_1984,chirinko_germanbanks_2006}, managers can
improve their power by buying other firms instead of paying dividends
\cite{jensen_takeovers_1986}, speculation on stock prices as well as
dividend earnings can be a significant source of
revenue~\cite{modiglia_cost_1958,
porta_dividend_2000,jensen_takeovers_1986}, and companies can have
strategic advantages, e.g.\ due to knowledge
sharing~\cite{teece_cooperation_1992,hamel_strategic_1991,dyer_relational_1998}.
Another possible hypothesis for control centralization is that managers
collude to form influential alliances: Indeed, agents (e.g. board
members) often work for different firms in central
positions~\cite{battiston_boards_2004}. Although all these factors are
likely to play a role, we here investigate a different hypothesis,
namely that a centralized structure may arise spontaneously, as a result
of a simple ``richt-get-richer'' dynamics~\cite{simon_richer_1955},
without any explicit underlying strategy from the part of the
companies. We consider a simple adaptive feedback
mechanism~\cite{gross_adaptive_2008}, which incorporates the indirect
control that companies have on other companies they own, which in turn
increases their buying power. The higher buying power can then be used
to buy portions of more important companies, or a larger number of less
important ones, which further increases their relative control, and
progressively marginalizes smaller companies. We show that this simple
dynamical ingredient suffices to reproduce many of the qualitative
features observed in the real data~\cite{vitali_network_2011}, including
the emergence of a core-periphery structure and the relative portion of
control exerted by the dominating core. Although this does not preclude
the possibility that companies may take advantage and further
consolidate their privileged positions in the network, it does suggest
that deliberate strategizing may not be the dominating factor which
leads to global centralization.

\section{Model description}\label{sec:model}

We consider a network of $N$ companies, where a directed edge between
two nodes $j\to i$ means company $j$ owns a portion of company $i$. The
relative amount of $i$ which $j$ owns is given by the matrix $w_{ij}$
(i.e. the ownership shares), such that $\sum_j w_{ij} = 1$. We note that
it is possible for self-loops to exist, i.e. a company can in principle
buy its own shares.  In the following, we describe a model with two main
mechanisms: 1. The evolution of the relative control of companies, given
a static network;
2. The evolution of the network topology via adaptive rewiring of the
edges.

\subsection{Evolution of control}

\begin{figure}[htb]
\begin{center}
  \includegraphics[width=.35\columnwidth]{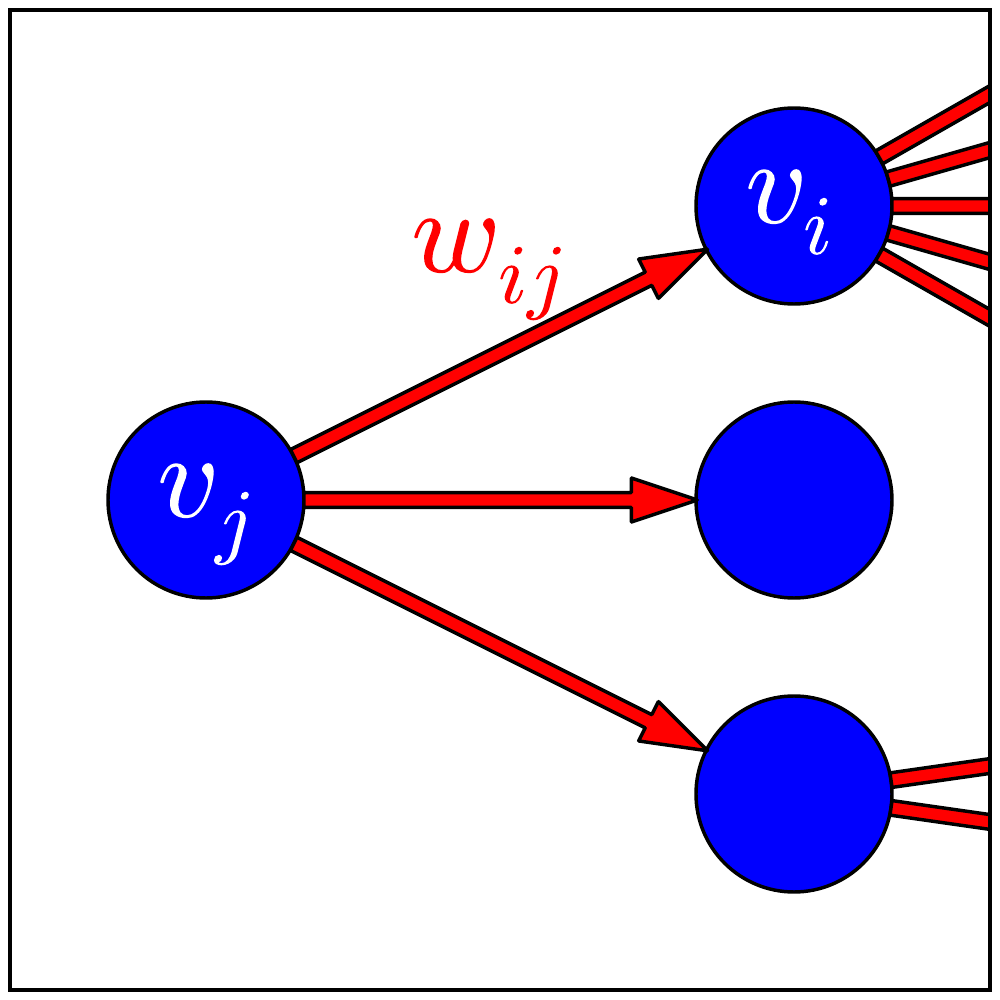}
  \includegraphics[width=.35\columnwidth]{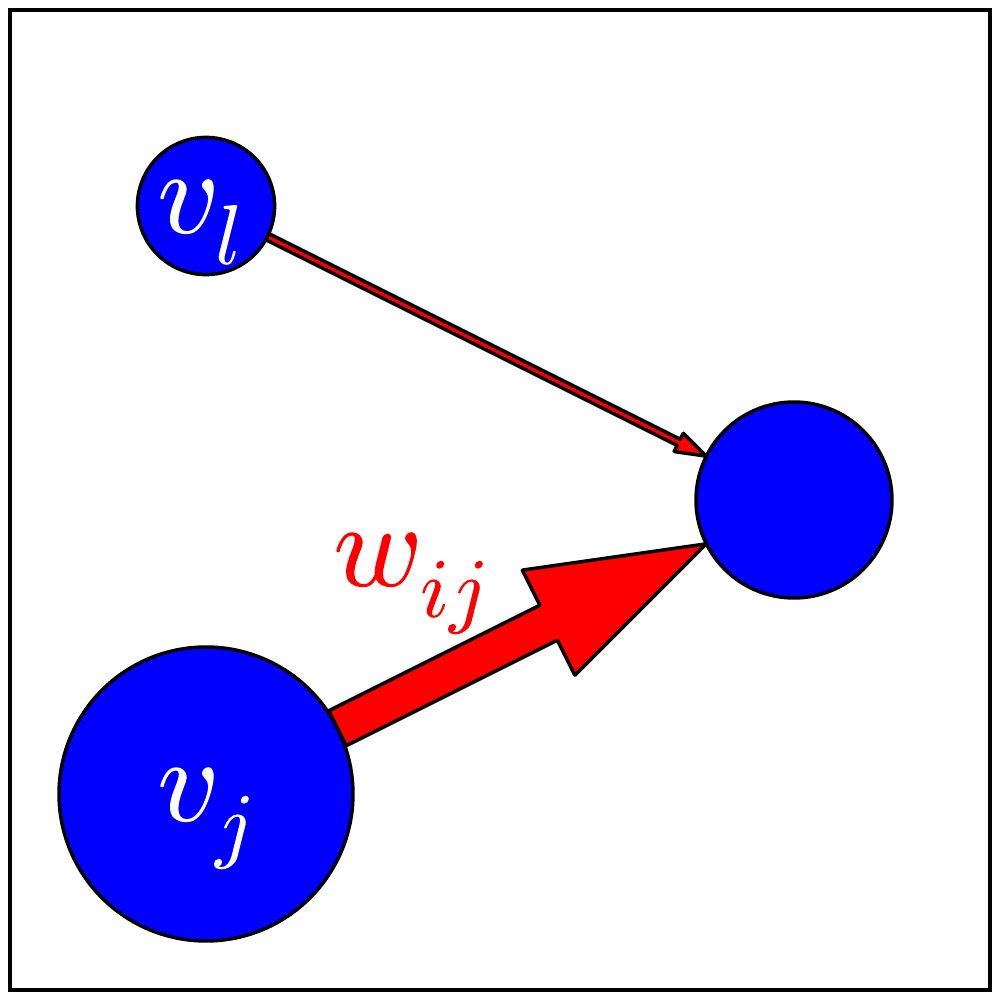}

  \caption{Illustration of the control of firms including indirect
  control (left) and the ownership being proportional to the control
  (right), as described in the text.}
  \label{fig:control}
\end{center}
\end{figure}

Here we assume that if $j$ owns $i$, it exerts some influence on $i$ in
a manner which is proportional to $w_{ij}$. If we let $v_j$ describe the
relative amount of control a company $j$ has on other companies, we can
write
\begin{equation}\label{eq:v}
  v_j = 1-\alpha + \alpha \sum_i A_{ij} w_{ij} v_i,
\end{equation}
where $A_{ij}$ is the adjacency matrix, the parameter $\alpha$
determines the propagation of control and $1-\alpha$ is an intrinsic
amount of independence between companies~\footnote{Eq.~\ref{eq:v} can be
seen as a weighted version of the Katz centrality
index~\cite{katz_new_1953}, which is one of many ways of measuring the
relative centrality of nodes in a directed network, such as
PageRank~\cite{page_PageRank_1999} and
HITS~\cite{kleinberg_authoritative_1999}. It converges for $0\leq\alpha<
1$ and we enforce normalization with $\sum_i v_i=N$.}. We further assume
that the control value $v_j$ directly affects other features such as
profit margins, and overall market influence, such that the buying power
of companies with large $v_j$ is also increased. This means that the
ownership of a company $i$ is distributed among the owners $j$,
proportionally to their control $v_j$, i.e.
\begin{equation}\label{eq:w}
  w_{ij} = \frac{A_{ij} v_j}{\sum_l A_{il} v_l},
\end{equation}
(see Fig.~\ref{fig:control}).
These equations are assumed to evolve in a faster time scale, such that
equilibrium is reached before the topology changes, as described in the
next section.

\subsection{Evolution of the network topology}

\begin{figure}[htb]
\begin{center}
  \includegraphics[width=.35\columnwidth]{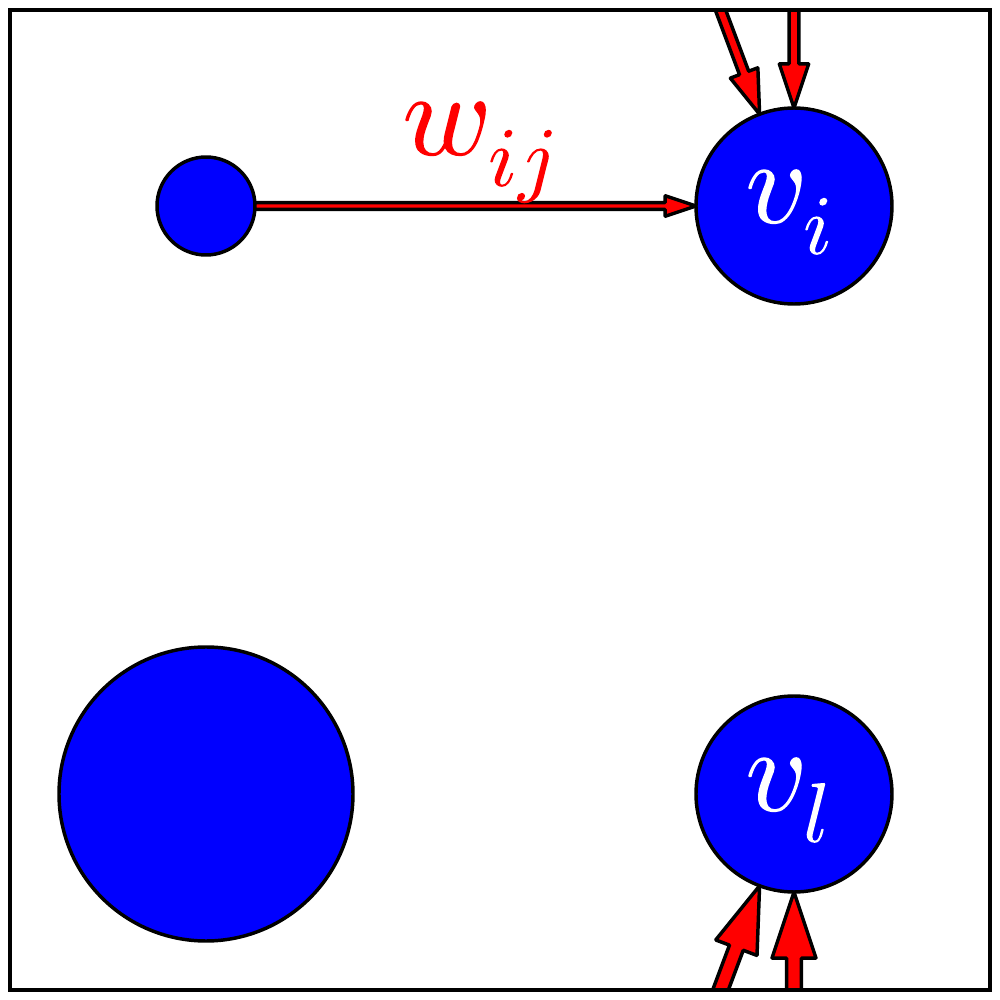}
  \includegraphics[width=.35\columnwidth]{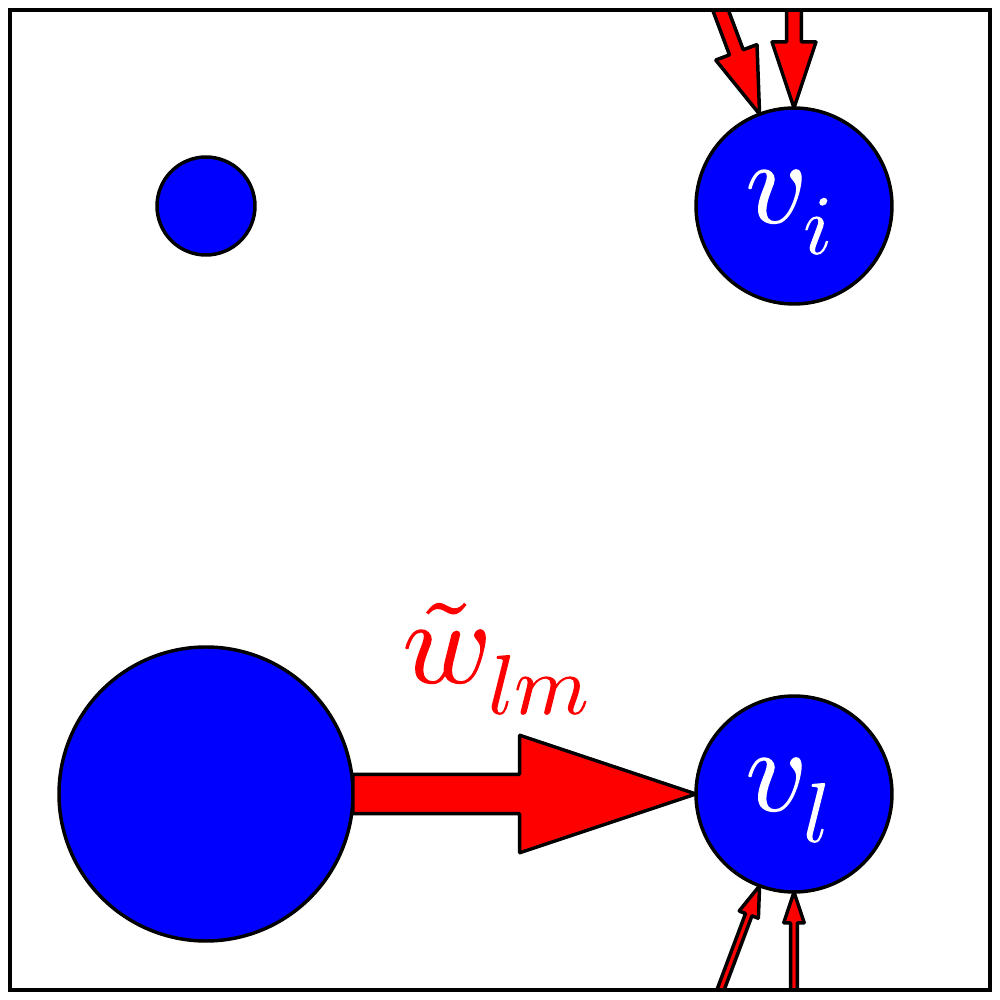}

  \caption{Illustration of the adaptive process, before the rewiring
  (left) and afterwards (right), as described in the text.}
  \label{fig:adaptiv}
\end{center}
\end{figure}

Companies may decide to buy or sell shares of a given company at a given
time. The actual mechanisms regulating these decisions are in general
complicated and largely unknown, since they may involve speculation,
actual market value, and other factors, which we do not attempt to model
in detail here. Instead, we describe these changes probabilistically,
where an edge may be deleted or inserted randomly in the network, and
such moves may be accepted or rejected depending on how much it changes
the control of the nodes involved.  For simplicity, we force the total
amount of edges in the network to be kept constant, such that a random
edge deletion is always accompanied by a random edge insertion. Such
``moves'' may be rejected or accepted, based on the change they bring to
the $v_j$ values of the companies involved. If we let $m$ be the company
which buys new shares of company $l$, and $j$ which sells shares of
company $i$, the probability that the move is accepted is
\begin{equation}\label{eq:rewire}
  p = \min\left(1,e^{\beta(\tilde{w}_{lm} v_l - w_{ij} v_i)}\right),
\end{equation}
where $w_{ij}$ is computed before the move and $\tilde{w}_{lm}$
afterwards, and the parameter $\beta$ determines the capacity companies
have to foresee the advantage of the move, such that for $\beta=0$ all
random moves are accepted, and for $\beta\to\infty$ they are only
accepted if the net gain is positive (see Fig.~\ref{fig:adaptiv}). Note
that in Eq.~\ref{eq:rewire} it is implied that companies with larger
control will tend to buy more than companies with smaller control, which
is well justified by our assumption that control is correlated with
profit and wealth.

The overall dynamics is composed by performing many rewiring steps as
described above, until an equilibrium is reached, i.e. the observed
network properties do not change any longer. In order to preserve a
separation of time scales between the control and rewiring dynamics, we
performed a sufficiently large number of iterations of Eqs.~\ref{eq:v}
and~\ref{eq:w} before each attempted edge move.


\section{Centralization of control}\label{sec:concentration}

\begin{figure*}[Htb!]
\begin{center}
  \begin{minipage}{0.3\textwidth}\centering
    \includegraphics[width=\textwidth]{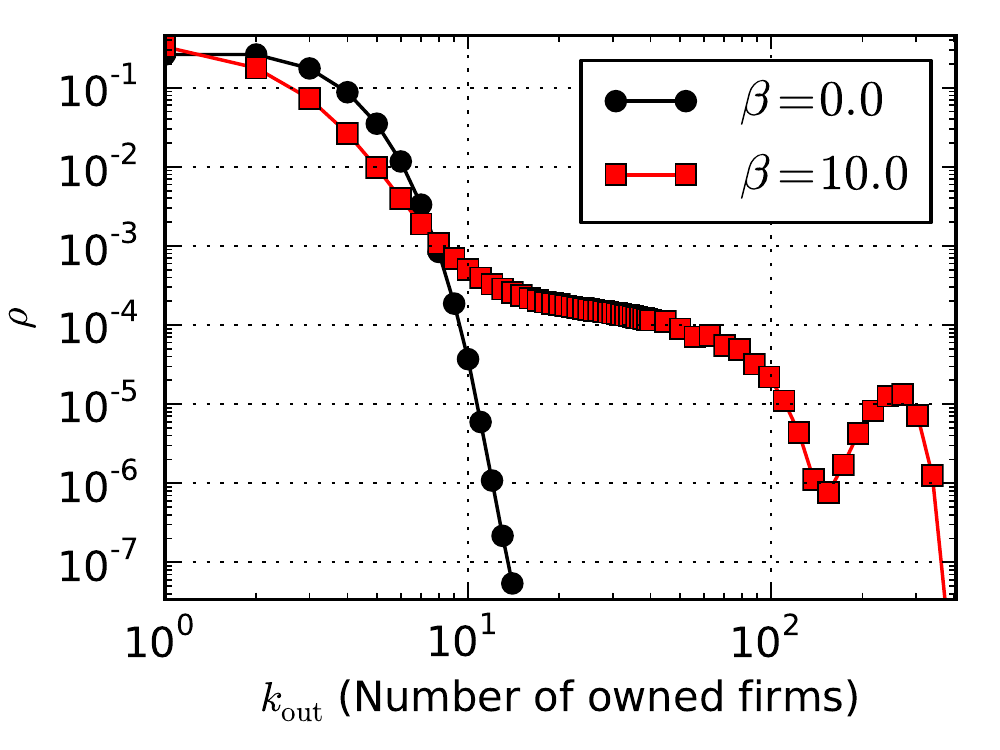}
    (a)
  \end{minipage}
  \begin{minipage}{0.3\textwidth}\centering
    \includegraphics[width=\textwidth]{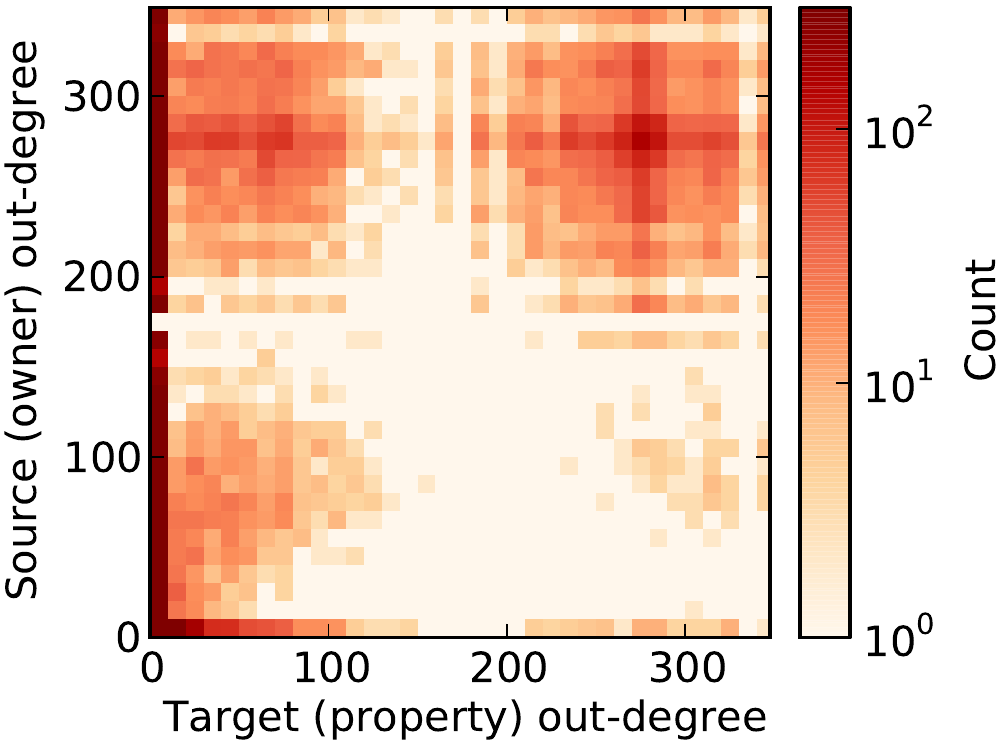}
    (b)
  \end{minipage}
  \begin{minipage}{0.24\textwidth}\centering
    \includegraphics[width=1\textwidth]{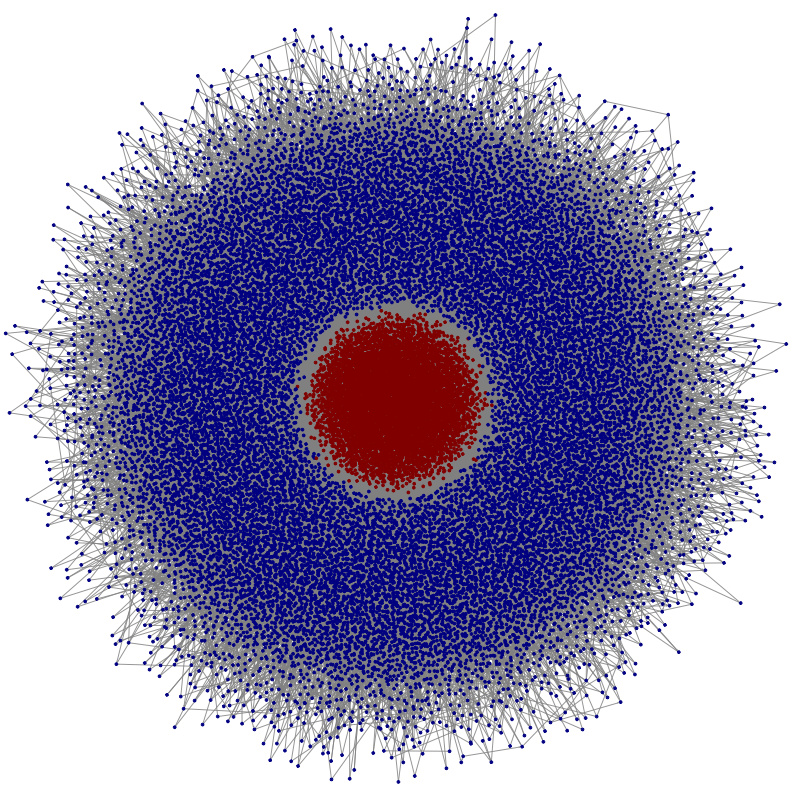}
    (c)
  \end{minipage}
  \begin{minipage}{0.14\textwidth}\centering
    \includegraphics[width=\textwidth]{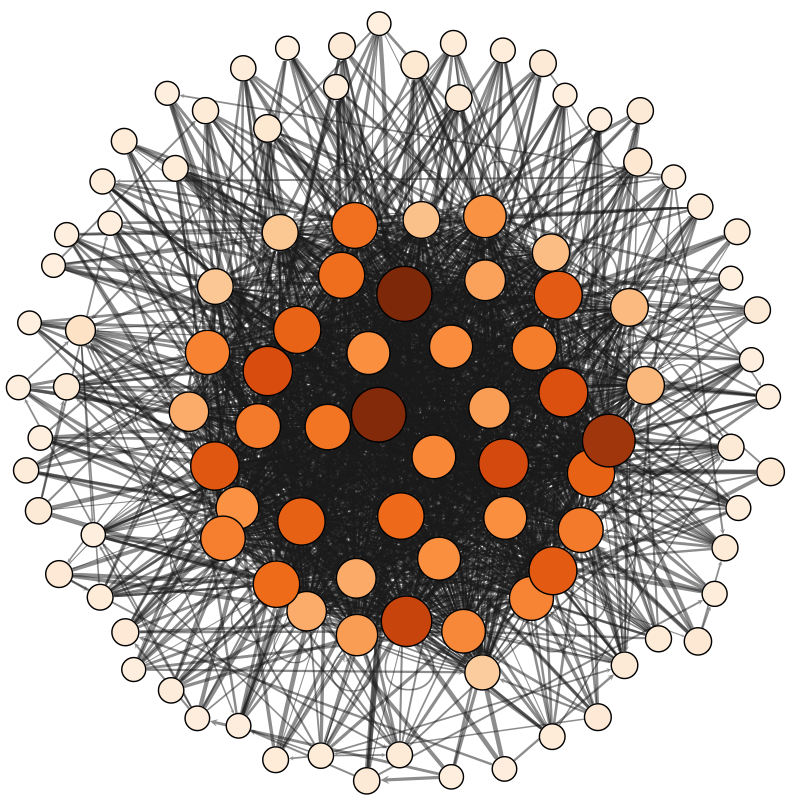}
    (d)
  \end{minipage}
\caption{(a) Degree distribution of the resulting network for
  $\left<k\right>=2$, a control propagation value of $\alpha=0.5$, 
  $N=30000$ and different values of prior knowledge $\beta$; 
  (b) Degree correlation matrix for $\beta=10$,
  showing the resulting core-periphery structure; (c) Graph layout of
  the whole network, with red nodes representing a chosen fraction of
  the most highly connected core, and blue ones the periphery; (d)
  Subgraph of the most powerful companies with $v_i > 20$ (about
  100). The node colors and sizes correspond to the $v_i$ values.}
  \label{fig:condensation}
\end{center}
\end{figure*}

A typical outcome of the dynamics can be seen in
Fig.~\ref{fig:condensation} for a network with $N=3\times10^4$ nodes and
average degree $\left<k\right>=2$, after an equilibration time of about
$6\times10^{9}$ steps.  In contrast to the case with $\beta=0$, which
results in a fully random graph, for a sufficiently high value of
$\beta$ the distribution of firm ownerships (i.e. the out-degree of the
nodes) becomes very skewed, with a bimodal form. We can divide the most
powerful companies into a broad range which owns shares from $10$ to
about $150$ other companies, and a separate group with
$k_{\text{out}}>150$. The correlation matrix of this network shows that
these high-degree nodes are connected strongly among themselves, and own
a large portion of the remaining companies (see
Fig.~\ref{fig:condensation}). This corresponds to a highly connected
``core'' of about 45 nodes with $\left<k_{\text{sub}}\right>\approx
39.8$, which is highlighted in red in Fig.~\ref{fig:condensation}c and
can be seen separately in Fig.~\ref{fig:condensation}d. The distribution
of in-degree (not shown) is bimodal as well with highest values for the
inner core. With values up to $k_{\rm in}=50$, the highest in-degree
(number of owners) is considerably below the highest out-degree (number
of firms owned at once).

\begin{figure}[htb]
\begin{center}
  \includegraphics[width=.49\columnwidth]{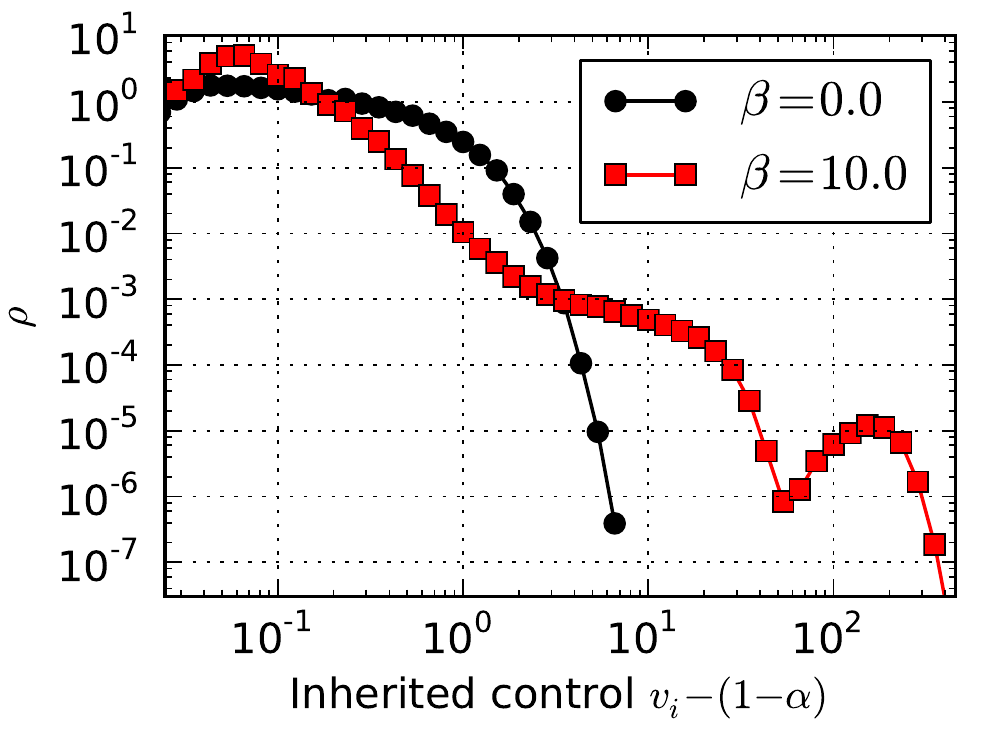}
  \includegraphics[width=.49\columnwidth]{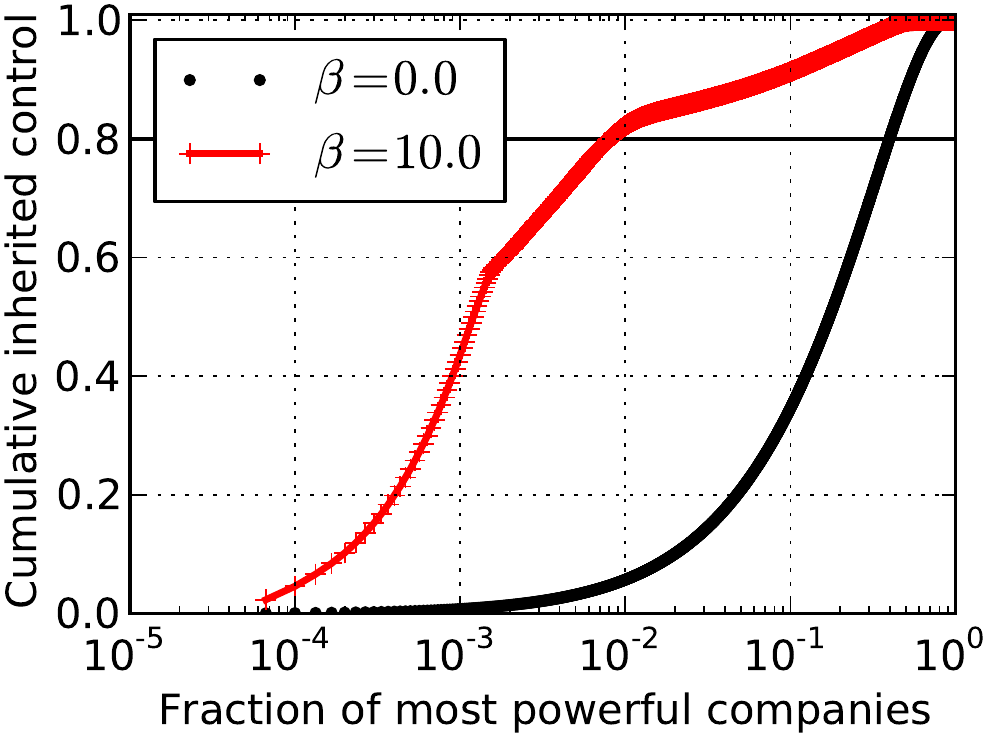}

  \caption{Left: Distribution of inherited control $v_i - (1-\alpha)$ for 
    $\alpha = 0.5$ and different values of $\beta$; Right: Relative fraction 
    of control as a function of fraction of most powerful companies.}
  \label{fig:condensation_v}
\end{center}
\end{figure}

Similarly to the out-degree, the distribution of control values $v_i$ is
also bimodal for larger values of $\beta$, as can be seen in
Fig.~\ref{fig:condensation_v}, and is strongly correlated with the
out-degree values. The total fraction of companies controlled by the
most powerful ones is very large, as shown on the right panel of
Fig.~\ref{fig:condensation_v}. For instance, we see that a fraction of
around $0.15\%$ of the central core controls about $57\%$ of all
companies. The companies with intermediary values of control (and
out-degree) also possess a significant part of the global control, e.g.
around $.85\%$ of the most powerful control an additional $25\%$ of the
network. It is important to emphasize the difference between these two
classes of companies for two reasons: Firstly the inner core inherits
control from intermediate companies without the need to gather up all
the minor companies.  In fact the ownership links going out from the
inner core (about $10^4$) is enough to cover the direct control of only
a third of all companies, while the effective control is more than a
half.  Secondly, the fraction of intermediary companies increases for
larger networks. For a network with $N=3\times10^5$, the inner core
includes a fraction of only $0.04\%$, controlling an effective $41\%$ of
the total companies. Nonetheless, all the most powerful companies
together account for around $1\%$ of the network and $82\%$ of the total
control; values which do not change considerably with system size.

Let us compare the results presented so far with empirical data
presented in \cite{vitali_network_2011}. For different reasons, this
comparison can only be qualitative. First of all, the empirical data
includes economic agents with different functions (shareholders,
transnational companies and participated companies) out of different
sectors (eg. financial and real economy), while we consider identical
agents.  Secondly, we force every company to be owned 100\%, while the
empirical data neglects restrained shares and diversified holdings.
Thirdly, the control analysis in \cite{vitali_network_2011} is done
somewhat differently: All the $600,508$ economic agents were considered
for the topological characterization, while many companies (80\% of all
agents there) were neglected for the control analysis. In the empirical
data, a strongly connected component of $1,318$ companies controls more
than a half of all companies arranged in the out component.  This
concentration is compatible with the core-periphery structure presented
in Fig.~\ref{fig:condensation}, however the empirical data does not show
a distinct bimodal structure. Nonetheless, there are highly connected
substructures in the core, e.g. a structure with 22 highly connected
financial companies ($\left<k_{\rm
sub}\right>\approx 12$) was highlighted in
\cite{battiston_debtrank_2012}.  The control concentration in the
empirical data was reported as a fraction of $0.5\%$ which controls
$80\%$ of the network. This is similar to the results of our model (see
Fig.~\ref{fig:condensation_v} on the right). There are, however,
features that our model does not reproduce, the most important of which
being the out-degree distribution of the network, which
in~\cite{vitali_network_2011} is very broad, and displays no discernible
scales, where in our case it is either bimodal or Poisson-like. One
possible explanation for this discrepancy is that we have focused on
equilibrium steady-state configurations of the dynamics, whereas the
real economy is surely far away from such an equilibrium. A more precise
model would need to incorporate such transient dynamics in a more
realistic way. Nevertheless, the general tendency of the control to be
concentrated on relatively few companies is evident in such equilibrium
states, and features very prominently in the empirical data as well.

\subsection{Transition to centralization}

\begin{figure}[htb]
\begin{center}
  \includegraphics[width=0.49\columnwidth]{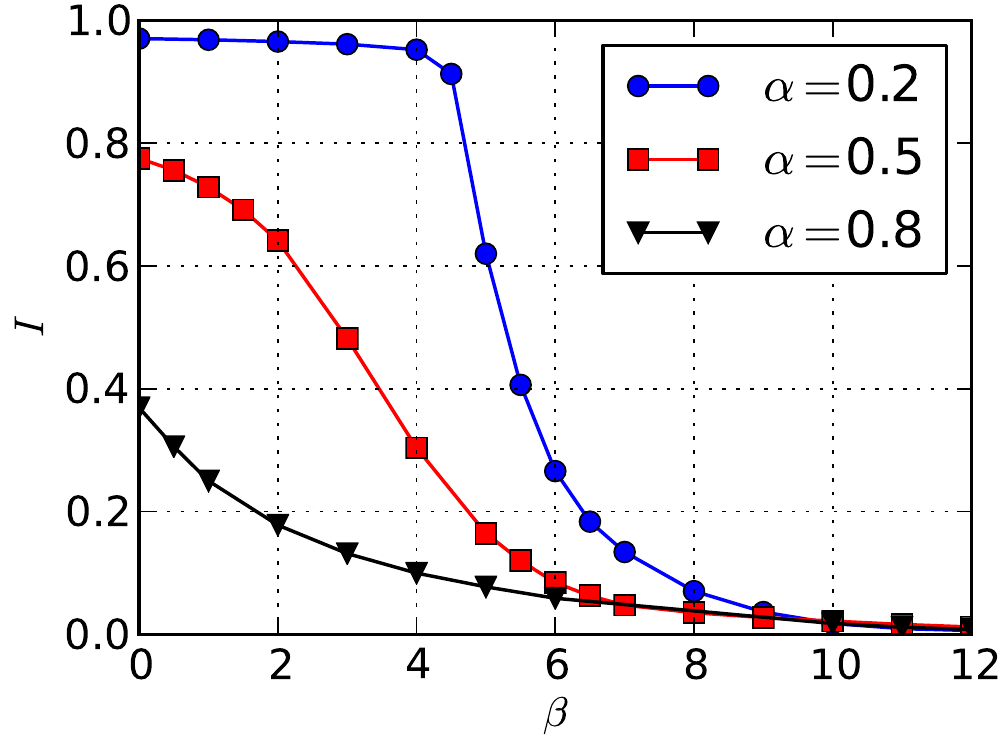}
  \includegraphics[width=0.49\columnwidth]{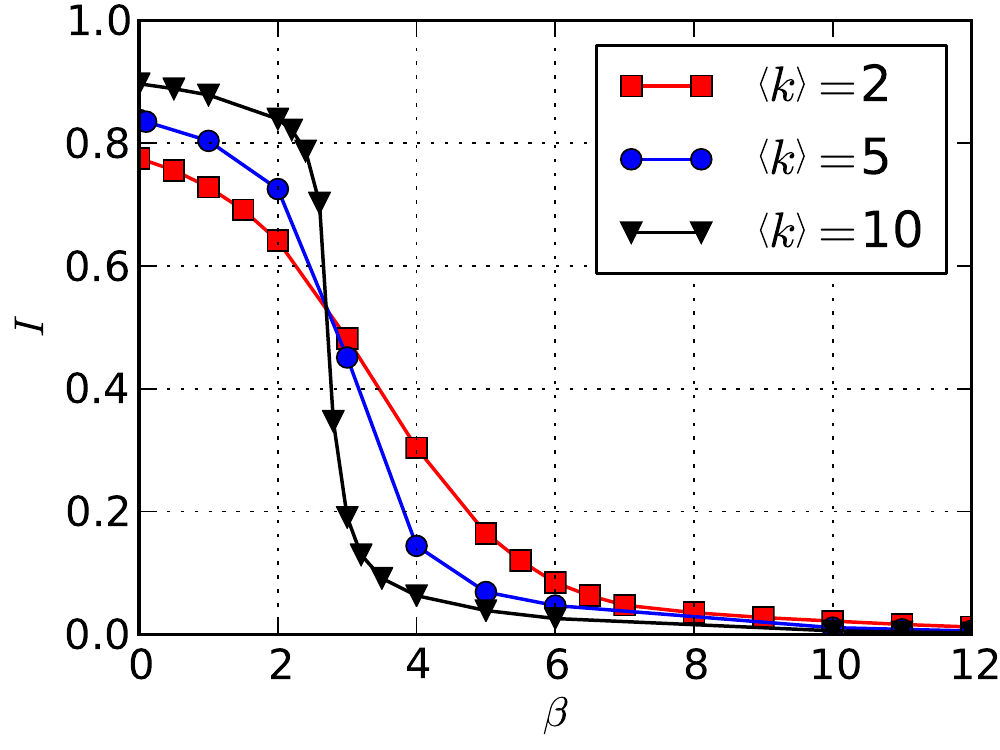}

  \caption{Inverse participation ratio $I=\left[\frac{1}{TN}\sum_{ti}
  v_i(t)^2 \right]^{-1}$ as a function of $\beta$, for a network with
  $N=10^4$, and for (left) $\left<k\right>=2$ and different values of
  $\alpha$ and (right) $\alpha=0.5$ and different values of
  $\left<k\right>$.}
  \label{fig:condensation_trans}
\end{center}
\end{figure}

To investigate the transition from homogeneous no centralized networks
with increasing $\beta$, we measured the inverse participation ratio
$I=\left[\frac{1}{TN}\sum_{ti} v_i(t)^2 \right]^{-1}$ with the time $t$
summing over a sufficiently long time window of length $T$ after
equilibration. Since $\frac{1}{N}\leq I\leq 1$, we expect $I=1$ in the
perfectly homogeneous case where $v_i=1$ for all nodes, and
$I=\frac{1}{N}$ if only one node has $v_i > 0$, and the control is
maximally concentrated. As can be seen in
Fig.~\ref{fig:condensation_trans}, we observe a smooth transition from
very homogeneous companies connected in fully random manner for
$\beta=0$, to a pronounced concentration of control for increased
$\beta$, for which the aforementioned core-periphery is observed. The
transition becomes more abrupt when either the average degree
$\left<k\right>$ is increased or the parameter $\alpha$ (which
determines the fraction of inherited control) is decreased.

\begin{figure}[htb]
\begin{center}
  \includegraphics[width=0.49\columnwidth]{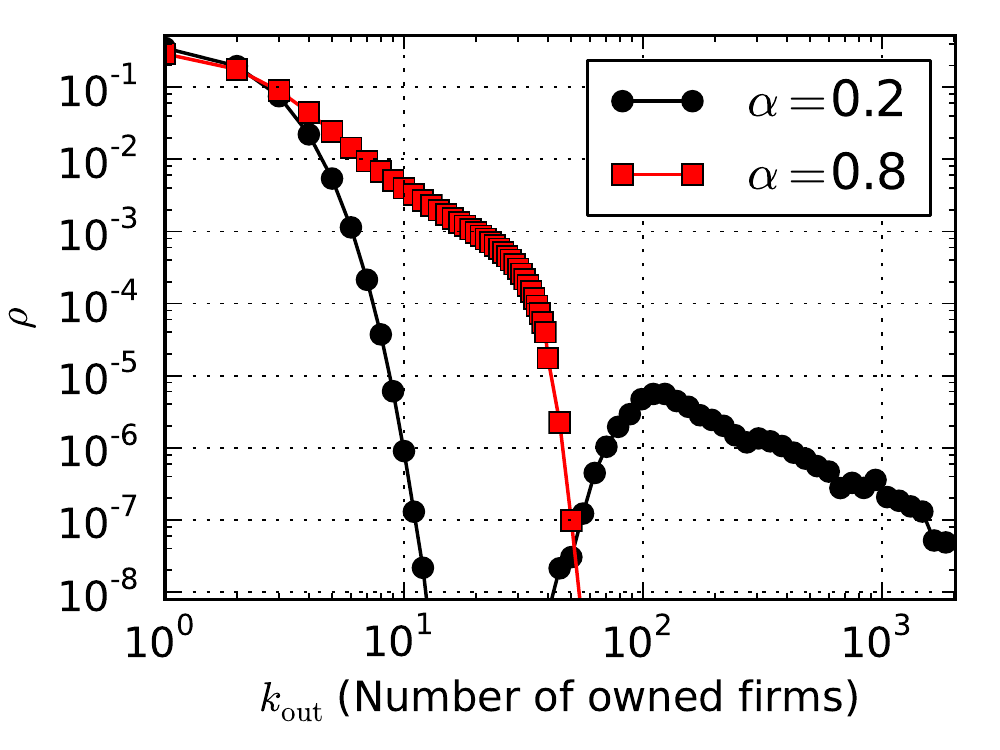}
  \includegraphics[width=0.49\columnwidth]{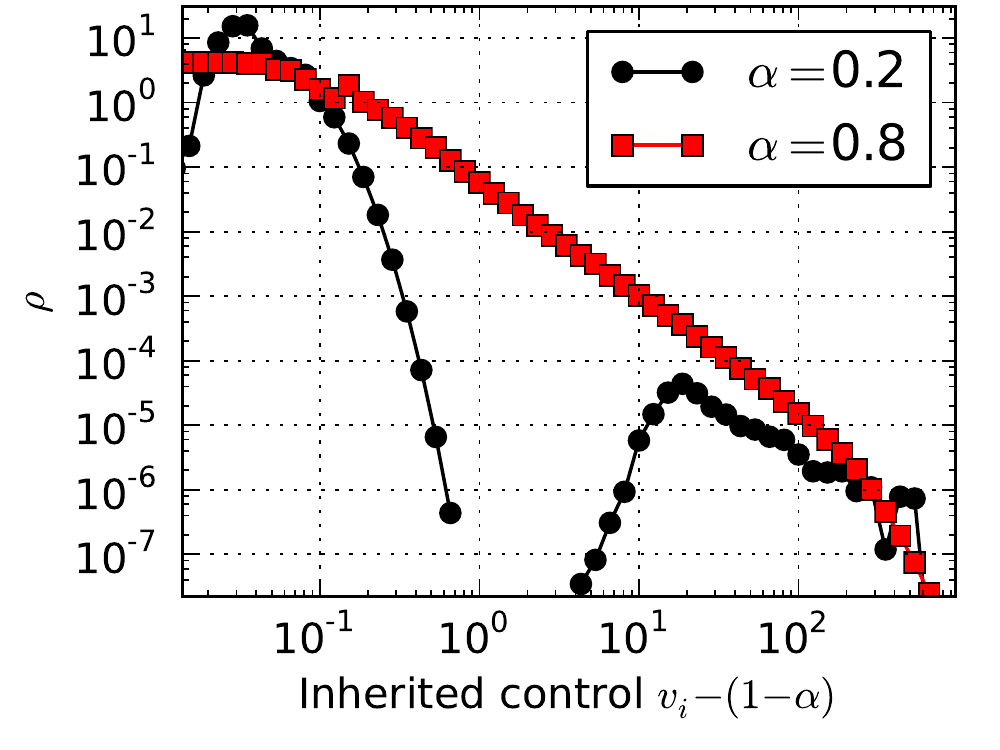}
  \caption{Distribution of out degrees (left) and inherited control 
  $v_i-(1-\alpha)$ (right) for $\beta=10$, $\left<k\right>=2$ and $N=30000$
  as in Fig.~\ref{fig:condensation} and \ref{fig:condensation_v}, 
  but for different values of $\alpha$.}
  \label{fig:condensation_dist}
\end{center}
\end{figure}

Centralization of control can emerge in different ways depending on the
parameters $\alpha$ and $\beta$. In Fig.~\ref{fig:condensation_dist}, it
is shown that different values of $\alpha$ for a high value of
$\beta=10$ can lead to a detached controlling core ($\alpha=0.2$) or to
broadly distributed control values ($\alpha=0.8$). With smaller values
of $\alpha$, indirect control is suppressed and companies can gain power
only by owning large numbers of marginal companies. E.g.:\ for
$\alpha=0.2$, this leads to a highly connected core of $41$ companies
having $\left<k_{\rm sub}\right>\approx 18.2$, the rest of the companies
have very little influence. For larger values of $\alpha$, indirect
control has a larger effect, which leads to a hierarchical network where
companies with small numbers of owned firms $k_{\rm out}$ may
nevertheless inherit large control values $v_i$.  The case with
$\alpha=0.5$ and $\beta=10$ shown in Figs.~\ref{fig:condensation} and
\ref{fig:condensation_v} exhibits a mixture of these two scenarios.  The
transition to a centralized core also occurs when increasing $\beta$ and
keeping $\alpha$ constant (see right panel in
Fig.~\ref{fig:condensation_trans}).

\begin{figure}[htb]
\begin{center}
  \includegraphics[width=0.41\columnwidth]{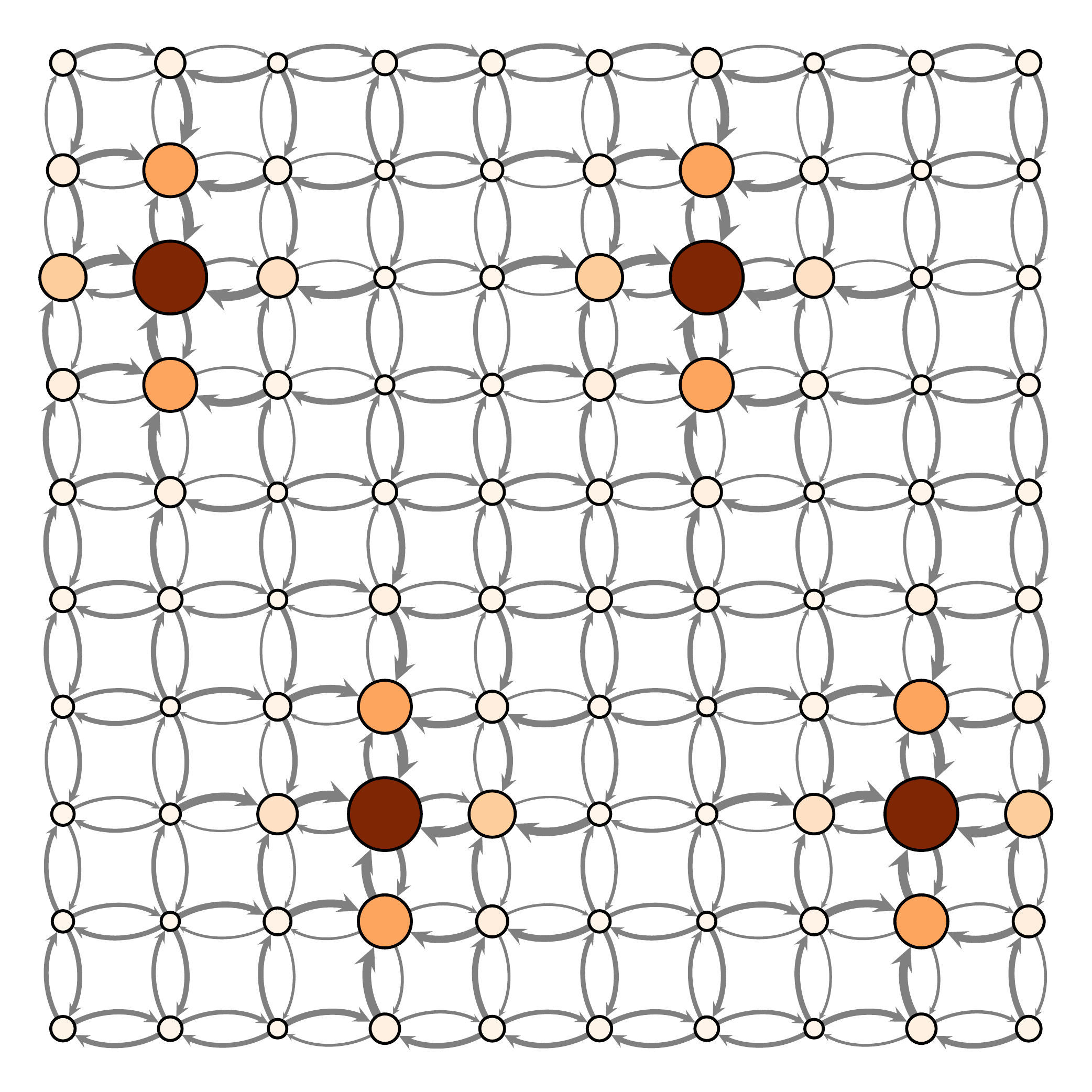} 
  \includegraphics[width=0.57\columnwidth]{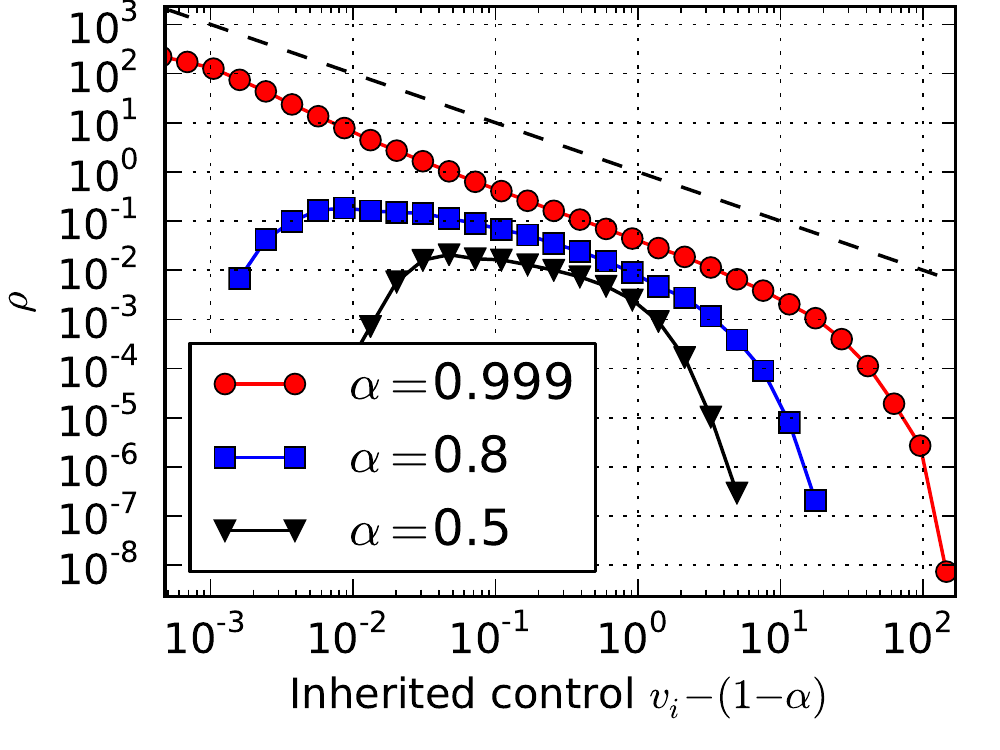}
  \caption{Left: Graph layout of a $10\times 10$ lattice with 
  $\alpha=0.9$. The vertex sizes and colors correspond to the $v_i$
  values, and the edge thickness to the $w_{ij}$ values.
  Right:
  Distribution of inherited control $v_i-(1-\alpha)$ for static poisson graphs 
  having $\left<k\right>=2$ and $N=30\,000$, with different values of
  $\alpha$ (for $\alpha=0.5$ and $\alpha=0.8$ shifted). 
  The dashed line is a power law with exponent $-1$.}
  \label{fig:lattice}
\end{center}
\end{figure}

One interesting aspect of the centralization of control as we have
formulated is that it is not entirely dependent on the adaptive
dynamics, and occurs also to some extent on graphs which are
static. Simply solving Eqs.~\ref{eq:v} and~\ref{eq:w} will lead to a
non-trivial distribution of control values $v_i$ which depend on the (in
this case fixed) network topology and the control inheritance parameter
$\alpha$. In Fig.~\ref{fig:lattice} is shown on the left the control
values obtained for a square $2D$ lattice with periodic boundary
conditions, and bidirectional edges. What is observed is a spontaneous
symmetry breaking, where despite the topological equivalence shared
between all nodes, a hierarchy of control is formed, which is not unique
and will vary between each realization of the dynamics. A similar
behavior is also observed for fully random graphs, as shown on the right
of Fig.~\ref{fig:lattice}, where the distribution of control values
becomes increasingly broader for larger values of $\alpha$,
asymptotically approaching a power-law $\rho(v) \sim v^{-1}$ for
$\alpha\to1$. This behavior is similar to a phase transition at
$\alpha=1$, where at this point Eq.~\ref{eq:v} no longer converges to a
solution.

\section{Conclusion}

We have tested the hypothesis that a rich-get-richer process using a
simple, adaptive dynamics is capable of explaining the phenomenon of
concentration of control observed in the empirical network of company
ownership~\cite{vitali_network_2011}. The process we proposed
incorporates the indirect control that companies have on other companies
they own, which increases their buying power in a feedback fashion, and
allows them to gain even more control. In our model, the system
spontaneously organizes into a steady-state comprised of a well-defined
core-periphery structure, which reproduces many qualitative observations
in the real data presented in \cite{vitali_network_2011}, such as the
relative portion of control exerted by the dominating companies.  Our
model shows that this kind of centralized structure can emerge without
it being an explicit goal of the companies involved. Instead, it can
emerge simply as a result of individual decisions based on local
knowledge only, with the effect that powerful companies can increase
their relative advantage even further.

It is interesting to compare our model to other agent based models
featuring agents competing for centrality. The emergence of
hierarchical, centralized states with interesting patterns of global
order was reported for agents creating links according to game theory
\cite{holme_centrality_2006,lee_multiadaptive_2011,do_patterns_2010} as
well as for very simple effective rules of rewiring according to
measured centrality
\cite{koenig_centrality_2011,bardoscia_climbing_2013}. The latter is
combined with phase transitions according to the noise in the rewiring
process. The stylized model of a society studied in
\cite{bardoscia_climbing_2013} shows a hierarchical structure, if the
individuals have a preference for social status. The intuitive emergence
of hierarchy is associated with shrinking mobility of single agents
within the hierarchy. This effect is present in our model as well and
deserves further investigation.

Our results may shed light on certain antitrust regulation strategies.
As we found that a simple mechanism without collusion suffices for
control centralization, any regulation which is targeted to diminish
such activities may prove fruitless. Instead, targeting the
self-organizing features which lead to such concentration, such as
e.g. limitations on the indirect control of shareholders representing
other companies, may appear more promising.

\bibliographystyle{apsrev4-1}
\bibliography{bib}

\end{document}